\documentclass[preprint,showpacs,preprintnumbers,amsmath,amssymb]{revtex4}

\usepackage{graphicx}
\usepackage{dcolumn}
\usepackage{bm}

\begin{document}

\preprint{APS}

\title{Finite-Size Scaling for Quantum Criticality \\
       above the Upper Critical Dimension:\\
       Superfluid-Mott-Insulator Transition in Three Dimensions}
\author{Yasuyuki Kato}
  \email{katoyasu@issp.u-tokyo.ac.jp}
\author{Naoki Kawashima}
  
\affiliation{%
  Institute for Solid State Physics, University of Tokyo,
  5-1-5 Kashiwa-no-ha, Kashiwa, Chiba 277-8581, Japan
}%

\date{\today}

\begin{abstract}
Validity of modified finite-size scaling above the upper critical dimension is demonstrated for the 
quantum phase transition whose dynamical critical exponent is $z=2$.
We consider the $N$-component Bose-Hubbard model,
which is exactly solvable and exhibits mean-field type critical phenomena in the large-$N$ limit.
The modified finite-size scaling holds exactly in that limit.
However, the usual procedure, taking the large system-size limit 
with fixed temperature, does not lead to the expected (and correct) mean-field critical behavior
due to the limited range of applicability of the finite-size scaling form.
By quantum Monte Carlo simulation, it is shown that the same holds in the case of $N=1$.
\end{abstract}

\pacs{67.25.dj,64.70.Tg,37.10.Jk,64.60.an}
\maketitle

\section{Introduction}
Since the quantum phase transition to Mott insulator from superfluid was observed in the optical
lattice system \cite{greiner2002}, this quantum critical phenomena has been one of hot topics. \cite{kato2008}
This system is effectively described by Bose-Hubbard (BH) Hamiltonian. \cite{jaksch1998}
The zero-temperature phase diagram of BH model
has been well investigated \cite{fisher1989,capogrosso2007,kawashima2009}.
There are phase transition points called multicritical points whose dynamical critical exponent is $z=1$
and line of the other type of phase transition called generic transition whose dynamical critical exponent 
is $z=2$ on the zero-temperature phase diagram.
In this paper, we consider the generic transition (i.e., $z=2$) in three-dimensional systems. 
The three dimension ($d=3$) is above the upper critical dimension $d_u=2$. 
Therefore, this phase transition is exactly classified and its critical exponents should be identical to those of the mean-field theory.
To estimate the locations of critical points quantitatively,
we frequently apply the finite-size scaling to the data of finite-size systems calculated using quantum Monte Carlo (QMC) method.

Above the upper critical dimension, the finite-size scaling (FSS) should be modified due to a dangerous irrelevant variable. \cite{binder1985}
In contrast to the conventional FSS below $d_u$, the modified finite-size scaling (MFSS) \cite{brankov} 
is not justified by renormalization group or scaling theories.
However, its validity has been demonstrated for
the five dimensional Ising model \cite{binder1985,luijten1999,jones2005}, $O(n)$ model \cite{singh1988}
and $\phi^4$ model in large-$N$ limit \cite{chen1998,luijten1999}.
For the quantum phase transition with $z=1$, below the upper critical dimension,
a simple application of the FSS
is trivially possible by identifying the inverse temperature $\beta$ as just an additional dimension.
Actually, to estimate the multicritical point quantitatively,
{\v S}makov and S{\o}rensen \cite{smakov2005} applied the FSS with the additional argument $\beta/L$
to the multicritical point in $d=2$ case where the system is below the upper critical dimension because $d+z<4$. 
For the quantum phase transition with $z\neq 1$, below the upper critical dimension,
the application of FSS is also possible with the additional argument $\beta/L^z$ instead of $\beta/L$
on the ground that the ratio between the correlation time $\xi_\tau$
and the correlation length $\xi$ to the $z$-th power is $\xi_\tau/ \xi^z = O(1)$. \cite{fisher1989} 
Zhao {\it et al.}, applied the FSS to the case $z=2$ and $d=2$,
which is just the upper critical dimension, 
and succeeded in estimating the phase boundary on the zero-temperature phase diagram of their model. \cite{zhao2008} 
The purpose of the present paper is to demonstrate the validity of the MFSS in the case where 
$d>d_u$ and $z\neq 1$, both by Monte Carlo simulation and by exact solutions.
We consider the case $z=2$, $d=3$, i.e., above the upper critical dimension.
It seems a natural extension to add the argument $\beta/L^2$ to the scaling function of MFSS. \cite{binder1989}
Namely, we assume that 
the singular part of the free energy $F_s$ has the scaling form,
\begin{eqnarray}
F_s\left(r,\eta,\beta,L \right)\sim \tilde{Y}_F\left(\delta L^{(d+2)/2},\eta L^{3(d+2)/4},\beta/L^2\right),
\label{eq:qmfss}
\end{eqnarray}
with a universal scaling function ${\tilde{Y}_F}$,
where the definition of the free energy is $F \equiv -\ln \Xi$ with the partition function $\Xi$,
$r$ indicates the coefficient of the term including square of the order parameter in 
the Hamiltonian 
(e.g., the chemical potential $\mu$ or the hopping amplitude $t$ in the model (\ref{eq:hamiltonian}) described below),
for $\delta$ indicates the difference from the quantum critical point 
(e.g., $\delta=r-r_c$),
and $\eta$ is the field inducing the order parameter.

The critical exponents for the finite temperature behavior at 
quantum critical point should be identical to those of mean-field theory,
e.g., $\chi \sim T^{-3/2}$ where $\chi$ is susceptibility. 
However, as shown in Sec. \ref{sec:N1FSS}, 
the exponents derived by the limit $L\rightarrow \infty$ of scaling form 
(e.g., $\chi \sim T^{-5/4}$) are different from those of the mean-field theory.
The reason of this apparent contradiction is that
the scaling form (\ref{eq:qmfss}) is valid only when $\beta/L^2 = O(1)$.
That is, we cannot infinitize $L$ in Eq. (\ref{eq:qmfss}) while keeping $\beta$ finite.
In this paper, we show that the application of MFSS to the $z=2$ quantum critical point 
is reliable, if the condition of validity is satisfied, just as well as the conventional FSS below the upper critical dimension.

In Sec. \ref{sec:model}, we define $N$-component BH model.
In Sec. \ref{sec:N1FSS}, we focus on the $N=1$ case and show the application of the MFSS to 
the numerical result of the QMC simulation.
In Sec. \ref{sec:NinfFSS}, we focus on the $N=\infty$ case, which is exactly solvable even for finite systems, 
to show that the susceptibility obeys the MFSS form under the condition $\beta/L^2 = O(1)$.
In Sec. \ref{sec:summary}, we give a discussion and summary of this paper.


\section{$N$-component Bose-Hubbard model}\label{sec:model}
We consider the $N$-component BH model on the hypercubic lattice whose Hamiltonian is described as 
\begin{eqnarray}
{\mathcal H}_N=-\frac{t}{Z}\sum_{\alpha=1}^{N}\sum_{\langle i,j \rangle}\left( 
b^{\dag}_{\alpha i} b_{\alpha j}+b_{\alpha i} b^{\dag}_{\alpha j}
\right)
-\mu \sum_{\alpha=1}^{N}\sum_{i} b^{\dag}_{\alpha i} b_{\alpha i}
+\frac{U}{2N} \sum_{\alpha=1}^{N}\sum_{\beta=1}^{N} \sum_{i} b^{\dag}_{\alpha i} b^{\dag}_{\beta i} b_{\beta i} b_{\alpha i},
\label{eq:hamiltonian}\end{eqnarray}
where $b^{\dag}_{\alpha i}$ ($b_{\alpha i}$) creates (annihilates) a $\alpha$-type boson at site $i$, 
and $\langle i,j \rangle$ runs over all pairs of nearest-neighbor sites.
The symbols $t$, $U$, and $\mu$, denote the hopping amplitude, the on-site interaction between bosons, and the chemical potential, respectively.
The coordination number in the hypercubic lattice is $Z=2d$. We take the lattice spacing as our unit of distance.
For concreteness, we consider only three-dimensional case in this paper. (i.e., $Z=6$.)
Generalization to arbitrary dimensions should be straightforward.

Here, we define the free energy $F_\eta$ as
\begin{eqnarray}
F_\eta &\equiv& - \frac{1}{N}\ln {\rm Tr}\left[ e^{-\beta \left( \mathcal{H}_N-\eta \mathcal{Q}\right)} \right],\\ 
\mathcal{Q}&\equiv&\sum_{\alpha=1}^{N}\sum_{i}\left( b^{\dag}_{\alpha i}+ b_{\alpha i}\right),
\end{eqnarray}
with the field $\eta$ inducing the order parameter.

The $N$-component BH model (\ref{eq:hamiltonian}) is solvable in the large-$N$ limit. 
In Sec. \ref{sec:NinfFSS}, we demonstrate that the MFSS scaling (\ref{eq:qmfss}) exactly
describes the asymptotic behavior of the model (\ref{eq:hamiltonian}) 
in the large-$N$ limit.
We note here that an exactly solvable model similar to the present one was investigated in the 1980s. 
\cite{zannetti1980,cesare1982}
The model was defined with Bose field operators in the continuous space.
In these papers, the authors discussed the critical behavior in the thermodynamic limit
near the quantum critical point.
As a result, the mean-field type criticality was confirmed above the upper critical dimension.
(e.g., $\chi \sim \delta^{-1}.$)

\section{numerical verification of modified finite-size scaling}\label{sec:N1FSS}
\begin{figure}
\includegraphics[trim=0mm -3mm 0mm 0mm,angle=0,scale=0.85]{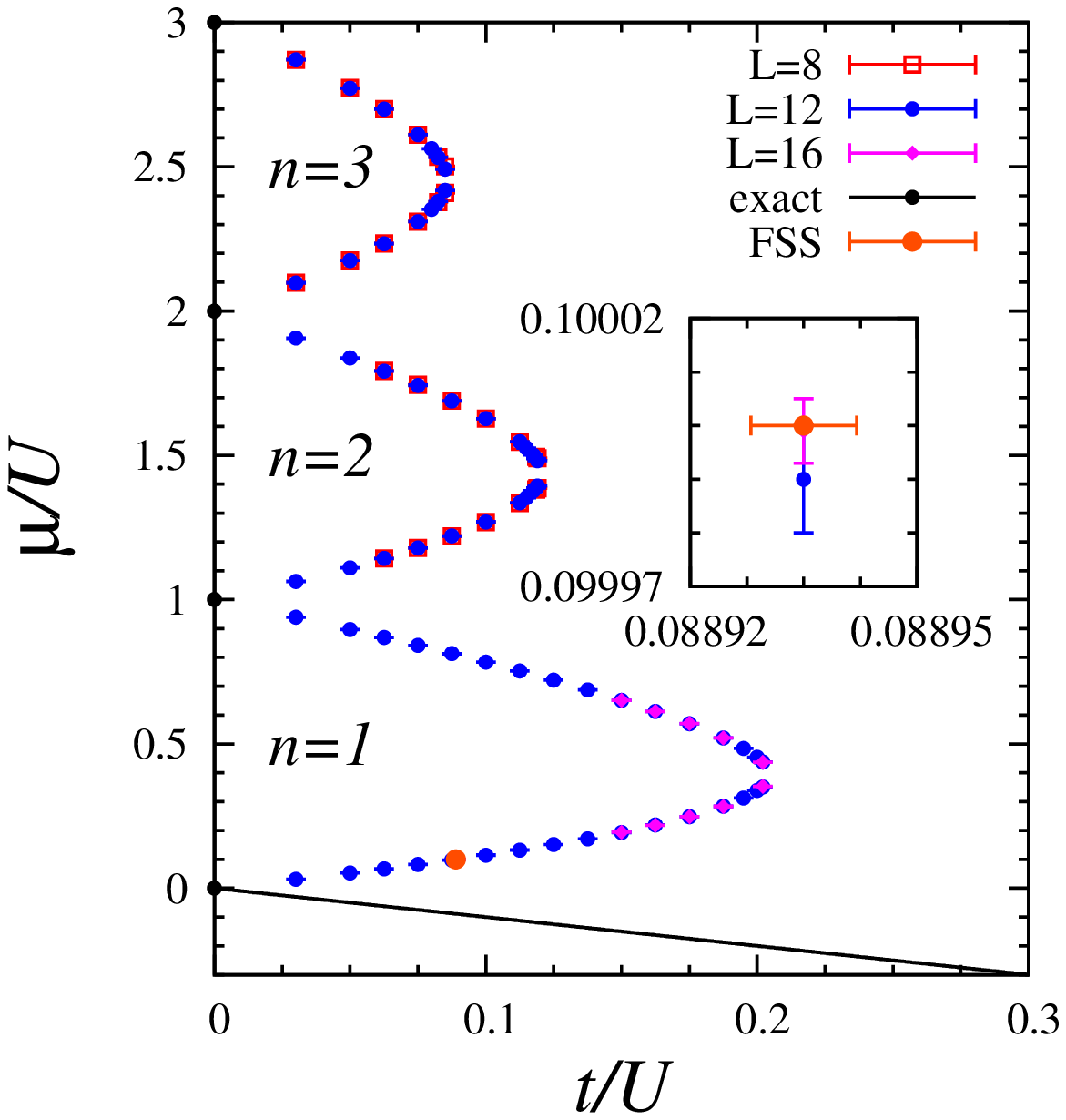}
\caption{\label{fig:compgapfss} (Color online)
  Zero-temperature phase diagram of single component BH model in 3D \cite{kawashima2009}.
  FSS indicates the result of MFSS.
}
\end{figure}

In this section, we apply the MFSS to the result of QMC simulation 
for the single component BH model \cite{kato2007,kato2009}.
We focus on the superfluid to Mott insulator transition.
The zero-temperature phase diagram is shown in Fig. \ref{fig:compgapfss}, 
which consists of Mott lobes and a superfluid region.
The phase boundary was estimated using the Mott gap. \cite{kawashima2009}
At the tip of the Mott lobe, which is a the multicritical point, 
the dynamical critical exponent $z$ is $1$ 
because of the asymptotic particle-hole symmetry. \cite{fisher1989,smakov2005}
The rest of critical lines corresponds to the generic transition with
the dynamical critical exponent $z=2$.
In this section, we fix the chemical potential as $\mu/U=0.1$ and 
vary the hopping amplitude $t/U$. 
Namely, $\delta$ in the first argument of the scaling functions corresponds to
$\delta=t/U-(t/U)_c$ in the present case.

\begin{figure}
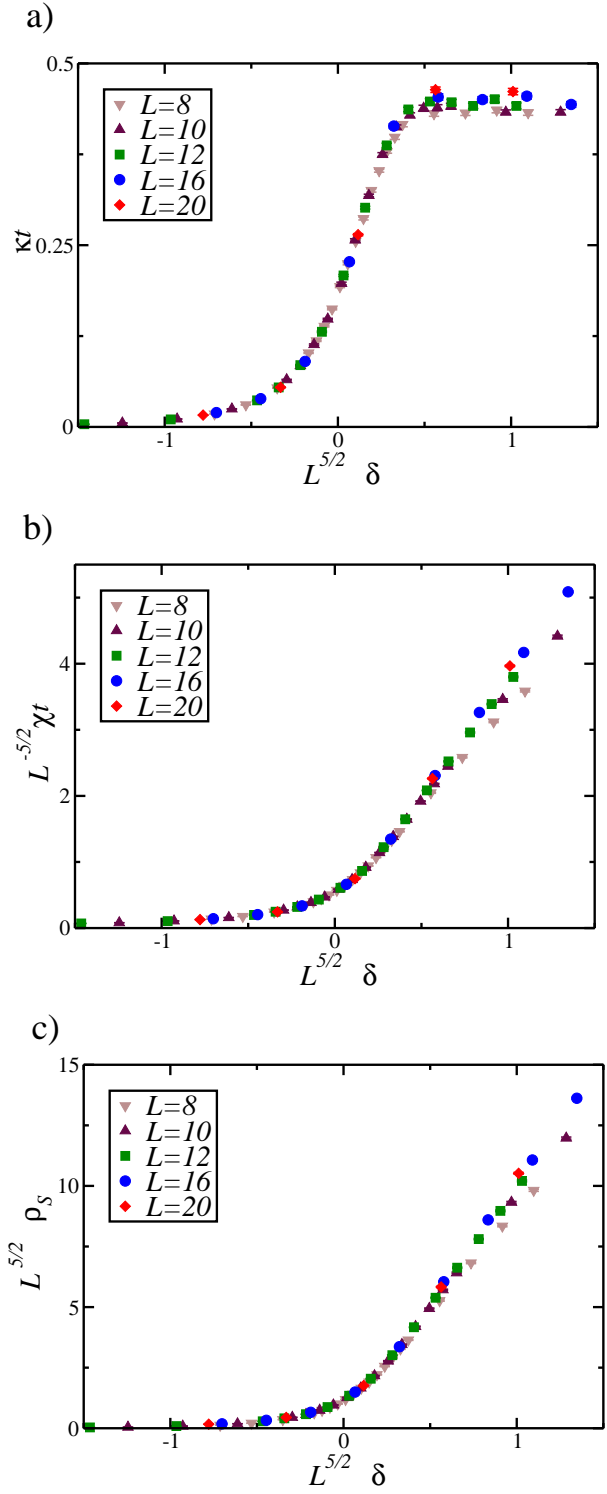

\includegraphics[trim=0mm -10mm 0mm 0mm,angle=0,scale=.32]{fsskappa.eps}
\hspace{15mm}
\includegraphics[trim=0mm -10mm 0mm 0mm,angle=0,scale=.32]{fsschi.eps}
\hspace{30mm}
\includegraphics[trim=0mm -10mm 0mm 0mm,angle=0,scale=.32]{fssrhos.eps}
\caption{\label{fig:qfss} (Color online) 
MFSS plots of the single component BH model where $\mu/U=0.1$, $\beta t/L^2 = 0.375$.
$\delta \equiv t/U - (t/U)_c$ with $(t/U)_c=0.088935$:
a) compressibility,
b) susceptibility 
c) superfluid density.
}
\end{figure}

We study compressibility $\kappa$ and susceptibility $\chi$.
Their definitions are
\begin{eqnarray}
\kappa &\equiv& \frac{1}{\rho^2}\frac{\partial \rho}{\partial \mu}, 
\end{eqnarray}
and
\begin{eqnarray}
\chi   &\equiv& -\left. \frac{1}{2L^d\beta}\frac{\partial^2 F_\eta}{\partial \eta^2}\right|_{\eta=0}, 
\end{eqnarray}
where
\begin{eqnarray}
\rho&\equiv&-\frac{1}{L^d\beta}\frac{\partial F_0}{\partial \mu}.
\end{eqnarray}
The scaling forms of $\kappa$ and $\chi$ are 
derived using the scaling form of the free energy (\ref{eq:qmfss}) as
\begin{eqnarray}
\kappa \sim {\tilde Y}_\kappa \left( x,y \right), 
\chi  \sim L^{{5}/{2}}{\tilde Y}_\chi \left( x,y \right),\label{eq:mfsscompchi} 
\end{eqnarray}
where
\begin{eqnarray}
x = \delta L^{{5}/{2}},
y = \frac{\beta t}{ L^2}.
\end{eqnarray}
We fix the second argument as $y=0.375$ and estimate the critical value of $t/U$ as $(t/U)_c=0.088935(7)$ 
at $\mu/U =0.1$ using the MFSS of $\kappa$ and $\chi$ as shown in Figs.\ref{fig:qfss} a) and b).
In these plots, we used the mean-field values for the exponents, leaving the critical value of $t/U$ as the only fitting parameter.

As long as $\beta t/L^2 =O(1)$, we can use the MFSS form just as well as we do 
in the conventional FSS for estimating the critical value of the relevant parameter ($t/U$ in the present case).
To compare between the estimation using Mott gap and MFSS,
we estimate the Mott gap at $t/U=0.088935$, $\mu/U=0.5$ and plot the corresponding points 
on the inset of Fig. \ref{fig:compgapfss}.
As we see in the figure, the agreement is very good.

\begin{figure}
\includegraphics[trim=0mm 0mm 0mm 0mm,angle=0,scale=.37]{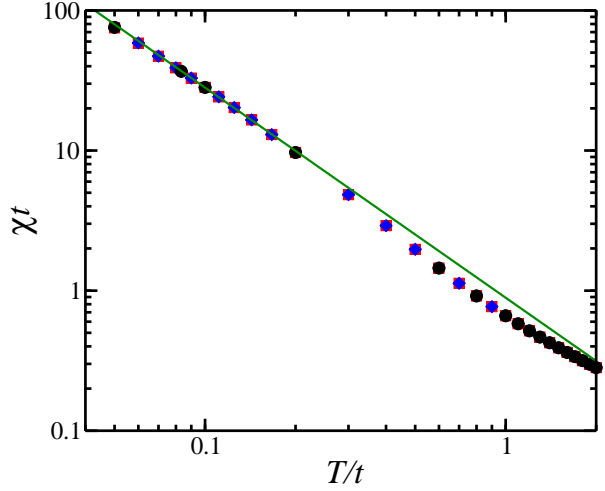}
\caption{\label{fig:chiATQCP} (Color online) 
  Temperature dependence of $\chi$ at the quantum critical point estimated by MFSS ($\mu/U=0.1$, $(t/U)_c=0.088935$).
  The solid line is $A(T/t)^{-3/2}$ where $A=0.89$.
  The data points are obtained for $L=24$, $32$, $48$.
  There is no visible size dependence on this scale.
  The statistical error is smaller than the symbol size.
}
\end{figure}

Here, a remark on the range of validity of the MFSS form is appropriate.
We consider the finite temperature behavior of $\chi$ at the quantum critical point $\delta=0$.
If we neglect the applicability condition of the MFSS form and
take the limit $L\rightarrow \infty$ while keeping $\beta t$ finite,
the finite temperature dependence of $\chi$ is derived as
\begin{eqnarray}
\chi &\sim& L^{5/2} \left( \frac{\beta t}{L^2}\right)^{5/4} \nonumber\\
     &\sim& T^{-5/4} {\rm(error !)},
\end{eqnarray}
from the scaling form (\ref{eq:mfsscompchi}).
This exponent $-5/4$ is different from that of mean-field theory $-3/2$.
As shown in Sec. \ref{sec:NinfFSS},
the reason of this error is that the scaling form (\ref{eq:mfsscompchi}) or (\ref{eq:qmfss}) is valid only under the condition of $\beta t/L^2 = O(1)$.
To confirm the mean-field exponent, we show the finite temperature dependence of $\chi$ at the quantum critical point in Fig. \ref{fig:chiATQCP}.

The superfluid density $\rho_S$ is one of the most important quantity characterizing the superfluidity.
However, it is not straightforward to derive the MFSS form of $\rho_S$ because
it is not directly obtained from the free energy by simple differentiation.
The superfluid density $\rho_S$ is proportional to the fluctuation of the winding number ${\bm W}=(W_x,W_y,W_z)$
and defined as $\rho_S \equiv \langle {\bm W}^2 \rangle / (\beta t L)$ within the framework of QMC simulation. \cite{pollock1987}
In Appendix \ref{app:fssrhos},
we show that $\rho_S=\chi/(\beta L^d)$, for the model (\ref{eq:hamiltonian})
in the large-$N$ limit under the condition, $\beta t/L^2 \geq O(1)$, $d>2$, and $\beta t \gg 1$.
From the MFSS for $\chi$, we obtain,
\begin{eqnarray}
\rho_S &\sim& L^{-\frac{5}{2}}{\tilde Y}_{\rho_S}\left( x,y \right).\label{eq:mfssrhos}
\end{eqnarray}
Although this form is derived only for the exactly solvable model,
we believe that this holds in general for the mean-field type critical behavior.
We apply this MFSS form to the result of $\rho_S$ estimated by QMC simulations.
As can be seen in Fig. \ref{fig:qfss} c), the MFSS (\ref{eq:mfssrhos}) describes the data well.

\section{Large-$N$ limit of $N$-component Bose-Hubbard Model}\label{sec:NinfFSS}
In this section, we consider the model (\ref{eq:hamiltonian}) that is known to exhibit a mean-field type
critical phenomena, to see if the MFSS is applicable to such a model.
We consider the model on the $d$ dimensional hypercubic lattice in the large-$N$ limit
and show that the MFSS form Eq. (\ref{eq:qmfss}) is exactly applicable to this case.
To derive the self-consistent equation of $\chi$ in the large-$N$ limit,
we represent the partition function as a functional integral
by making use of a coherent state basis at first.
Then, we use the Stratonovitch-Hubbard transformation and the saddle-point method,which is also called the steepest descent method.
Thus, the self consistent equation of susceptibility $\chi$ in large-$N$ limit is derived exactly as
\begin{eqnarray}
\chi^{-1}=-\mu-t+\frac{U}{L^d}\sum_{\bf k}\frac{1}{
\exp \left[  \beta\chi^{-1}+\frac{2\beta t}{Z} \sum_{\delta=1}^{d}\left( 1-\cos k_{\delta} \right) \right]-1}.
\end{eqnarray}
See Appendix \ref{app:detail_self} for details of the derivation.
By expanding the summand with respect to 
$\exp \left[  - \left\{ \beta\chi^{-1}+ \frac{2\beta t}{Z}\sum_{\delta=1}^{d}\left( 1-\cos k_{\delta} \right) \right\} \right]$,
we obtain
\begin{eqnarray}
\chi^{-1}=-\mu-t+\frac{U}{L^d}
\sum_{\nu=1}^{\infty} e^{-\nu \beta \chi^{-1}} \left[ 
\sum_{n=1}^{L}\exp \left[ -\frac{ 2\nu \beta t}{Z}\left\{ 1-\cos\left(\frac{2\pi n}{L}\right)\right\}\right]
\right]^d.
\label{eq:chi}
\end{eqnarray}
Below we show that this equation has a solution such that
$\chi\sim O((\beta t)^{(d+2)/4})$.
Therefore, we assume $\chi\sim O((\beta t)^{(d+2)/4})$ for $\chi$ in the r.h.s. of (\ref{eq:chi}).
Then, as shown in Appendix \ref{app:detail_appr}, the approximation formula
\begin{eqnarray}
\sum_{\nu=1}^{\infty} e^{-\nu \beta \chi^{-1}} \left[ 
\sum_{n=1}^{L}\exp \left[ -\frac{2\nu \beta t}{Z}  \left\{ 1-\cos\left(\frac{2\pi n}{L}\right)\right\}\right]
\right]^d \simeq \beta^{-1} \chi,
\label{eq:app}
\end{eqnarray}
becomes exact in the limit $\beta t \rightarrow \infty$ under the condition that
$d>2$, $\beta t/L^2 \ge O(1)$.
Using the self-consistent Eq. (\ref{eq:chi}) and the approximation (\ref{eq:app}),
we arrive at a simple equation
$\chi^{-1}=-\mu-t+U\chi\beta^{-1} L^{-d}$.
Its solution can be cast into the form,
\begin{eqnarray}
\frac{\chi t}{Z} &\simeq& \left( \frac{\beta t}{Z}\right)^{\frac{d+2}{4}} 
P_{\chi}^{UZ/t}\left( \left( \frac{\beta t}{Z}\right)^{\frac{d+2}{4}} Z\left(-\frac{\mu }{t}-1\right),
\left(\frac{\beta t}{L^2 Z}\right)^{-\frac{d}{2}}\right), \nonumber\\
&\simeq& L^{\frac{d+2}{2}} P_{\chi}^{UZ/t}\left( L^{\frac{d+2}{2}} Z\left(-\frac{\mu }{t}-1\right),\frac{\beta t}{L^2 Z}\right),
\label{eq:scaling_chi}
\end{eqnarray}
with a scaling function
\begin{eqnarray}
P_{\chi}^{u}\left(x,y\right) &\equiv& \frac{2}{x + \sqrt{x^2+4uy^{-1}}}.
\label{eq:scalingfchi}
\end{eqnarray}
At the critical point ($\mu=-t$), we obtain $\chi t \sim (\beta t)^{(d+2)/4} \times (\beta t / L^2)^{-d/4}$.
To make this consistent with $\chi = O((\beta t)^{(d+2)/4})$ assumed at first and the condition $\beta t/L^2 \ge O(1)$,
we must demand  $\beta t /L^2=O(1)$.
Thus, 
we have proved that Eq. (\ref{eq:chi}) has a solution $\chi=O((\beta t)^{(d+2)/4})$
that satisfies Eq. (\ref{eq:scaling_chi}), and
the MFSS form (\ref{eq:mfsscompchi}) has been derived as a formula that is asymptotically exact
under the condition of $d>2$ and $\beta t/L^2 = O(1)$.

\begin{figure}
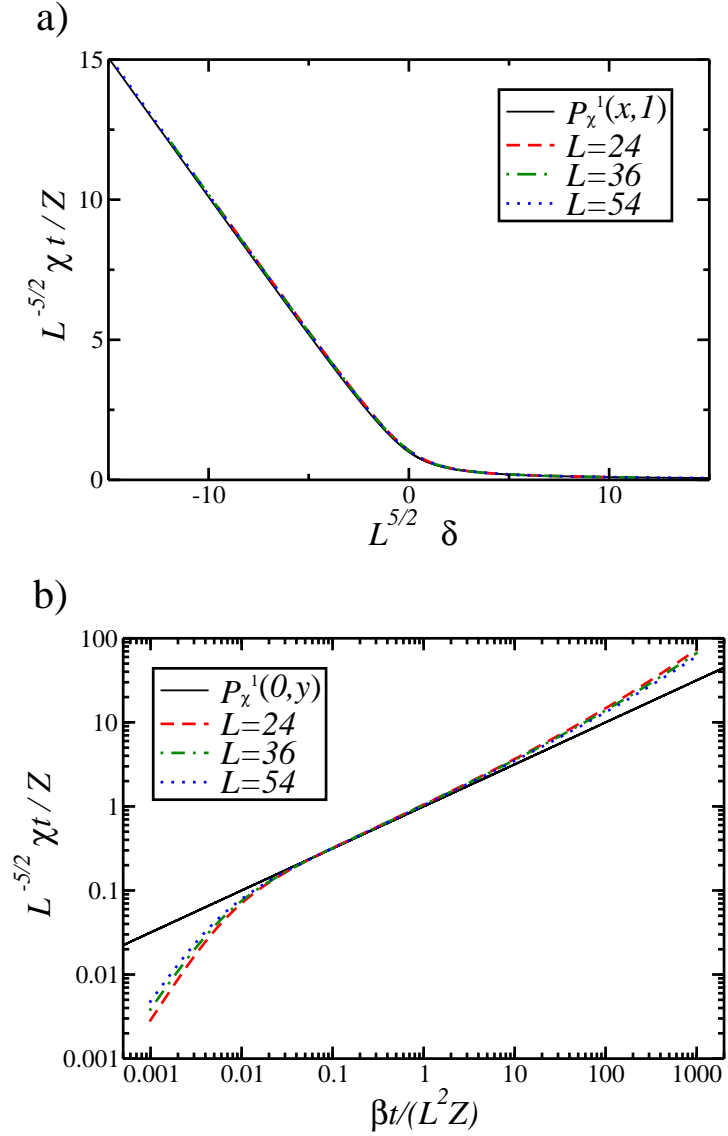

\includegraphics[trim=0mm -10mm 0mm 0mm,angle=0,scale=.37]{exactfsschi1.eps}
\includegraphics[trim=0mm 0mm 0mm 0mm,angle=0,scale=.37]{exactfsschi2.eps}
\caption{\label{fig:exactqfss} (Color online) 
MFSS plots of susceptibility at $U/(t/Z)=1$.
a) the $x$-dependence of scaling function of susceptibility $P^{UZ/t}_{\chi}(x,y)$ 
and the solution of self-consistent Eq.(\ref{eq:chi}) with $y\equiv\beta t/(ZL^2) =1$,
b) $y$-dependence of $P^{UZ/t}_{\chi}(x,y)$
and the solution of self-consistent Eq.(\ref{eq:chi}) at quantum critical point $\delta=0$. }
\end{figure}

To see the validity of the form of the scaling function (\ref{eq:scalingfchi}), 
we demonstrate the MFSS plot of susceptibility in $d=3$.
We solve the self-consistent Eq. (\ref{eq:chi}) without using Eq. (\ref{eq:app})
and plot on Figs. \ref{fig:exactqfss} a) and b).
As shown in Fig \ref{fig:exactqfss} b), the MFSS form fits well in the region $\beta t /L^2=O(1)$.


\section{Discussion and Summary}\label{sec:summary}
In Sec. \ref{sec:N1FSS} and \ref{sec:NinfFSS}, we have demonstrated that
the MFSS (\ref{eq:qmfss}) is efficient in locating quantum critical points
whose dynamical critical exponent is $z=2$.
It has been shown that the MFSS is valid only if 
the second argument of scaling function $\beta t/L^2$ is $O(1)$.
In particular, it is not permitted to infinitize $L$ in the scaling forms (\ref{eq:mfsscompchi}) and (\ref{eq:mfssrhos}) while keeping $\beta$ fixed.
This explains the apparent contradiction between the MFSS and the mean-field critical exponents.
It should be remarked here that
similar situations appear in classical models.
Suppose that we try to apply the MFSS to a finite-temperature phase transition 
of a classical system
and send the system size in some (not all) of the directions to infinity while keeping the size in other directions fixed. \cite{singh1986,binder1989}
Singh and Pathria \cite{singh1986} considered a system of size $L^{d-d'}\times {L'}^{d'}$ where $d$ is larger than the upper critical dimension
and $d'$ is less than the lower critical dimension.
They analyzed a spin model with $O(n)$ symmetry in the limit of $L' \rightarrow \infty$. \cite{singh1986}
Then they derived the scaling form of the susceptibility $\chi_0$ as
\begin{eqnarray}
\chi_0 \sim L^{\frac{2(d-d')}{4-d'}} Y^{d'}_{\chi_0}\left( \tilde{t} L^{\frac{2(d-d')}{4-d'}}\right),
\end{eqnarray}
where $\tilde{t}\equiv (T-T_c)/T_c$.
Then, $\chi_0 \sim L^{2(d-d')/(4-d')}$ at the critical point $\tilde{t}=0$. 
On the other hand, if we keep $L/L'$ is finite,
the MFSS form is
\begin{eqnarray}
\chi_0 \sim L^{\frac{d}{2}} {\overline Y}^{d'}_{\chi_0}\left( \tilde{t} L^{\frac{d}{2}}, L/L'\right),
\label{eq:mfssCL}\end{eqnarray}
with the additional argument $L/L'$. \cite{binder1989}
If we ignore the validity condition of the MFSS (\ref{eq:mfssCL}) and take the limit $L'\rightarrow \infty$,
we reach an erroneous conclusion, that is $\chi_0 \sim L^{d/2}$ at $\tilde{t}=0$.

In summary, the MFSS is applied to the quantum critical phenomena with the dynamical critical exponent $z=2$.
Using the $N$-component BH model, the MFSS form of the susceptibility Eq. (\ref{eq:scaling_chi})
is exactly derived in the large-$N$ limit with the applicability condition $d>2$ and $\beta t/L^2=O(1)$.
We also apply the MFSS to the numerical results obtained by QMC simulations.
As a result, we see that a position of quantum critical point estimated by MFSS is identical to 
that estimated by the Mott gap within the statistical error.
Finally, note that the scaling function derived in this paper $P^{u}_{\chi}(x,y)$ 
is in complete agreement with the scaling function of $\phi^4$ model derived in Ref. \cite{luijten1999}.
While the scaling function is not justified by the renormalization group or scaling theories
in contrast to the standard FSS below the upper critical dimension,
the agreement strongly indicates that 
the mean-field scaling function above the upper critical dimension is universal.


\begin{acknowledgments}
The present work was financially supported by 
a MEXT Grant-in-Aid for Scientific Research (B) (19340109), 
a MEXT Grant-in-Aid for Scientific Research on Priority Area ``Novel States of Matter Induced by Frustration'' (19052004), 
a Grant-in-Aid for JSPS Fellows,
and by the Next Generation Supercomputing Project, Nanoscience Program, MEXT, Japan.
The quantum Monte Carlo simulations were executed at the Supercomputer Center, Institute for Solid State Physics, University of Tokyo.
\end{acknowledgments}
\appendix
\section{Calculation of large N limit}\label{app:detail}
\subsection{Self-consistent equation of $\chi$}\label{app:detail_self}
Here, we derive the self-consistent equation (\ref{eq:chi}) of the Hamiltonian (\ref{eq:hamiltonian}).
The partition function is expressed as
\begin{eqnarray}
Z_{N} &=& \int \mathcal{D} {\bm \psi}_i(\tau) \mathcal{D} {\bm \psi}^{*}_i(\tau) e^{-\left(S_0+S_1\right)}, \nonumber \\
S_0   &=& \int_0^{\beta}d\tau \left[
\sum_{i} \left\{ 
{\bm \psi}_i^{*}(\tau) \cdot \left( \partial_{\tau}{\bm \psi}_i (\tau) \right)
-\mu {\bm \psi}_i^{*} (\tau) \cdot {\bm \psi}_i (\tau)
\right\} \right. \nonumber\\ && \left.
 -\frac{t}{Z}\sum_{\langle i,j \rangle}\left( 
 {\bm \psi}_i^{*} (\tau) \cdot {\bm \psi}_j (\tau)
+ {\bm \psi}_j^{*} (\tau) \cdot {\bm \psi}_i (\tau)
\right)\right],
\label{eq:partitionfunction}\\
S_1   &=& \int_0^{\beta}d\tau 
\frac{U}{2N}\sum_{i} \left( {\bm \psi}_i^{*} (\tau) \cdot {\bm \psi}_i (\tau) \right)^2,
 \nonumber
\end{eqnarray}
by the path-integral representation with ${\bm \psi}_i (\tau)$ being an $N$-component complex field.
Using Stratonovitch-Hubbard transformation, the partition function 
is written as
\begin{eqnarray}
Z_{N} &=& \int \mathcal{D} {\bm \psi}_i(\tau) \mathcal{D} {\bm \psi}^{*}_i(\tau) \mathcal{D} s_i(\tau)
e^{- S_0 }\nonumber\\&&
\times \exp \left[
-\int_0^{\beta}d\tau \left\{
\sum_i\left(
\frac{N}{2}s^2_i(\tau) -i\sqrt{U}s_i(\tau) \left( {\bm \psi}_i^{*} (\tau) \cdot {\bm \psi}_i (\tau)  \right)
\right)\right\}\right], \nonumber \\
&=& \int \mathcal{D} s_i(\tau) e^{-\frac{N}{2}\int_0^\beta d\tau \sum_i s_i^2(\tau)}\left[ Z_1\left(\left\{ s\right\}\right) \right]^N,\\
Z_1\left( \left\{ s \right\}\right) &\equiv&
\int \mathcal{D} \psi^{\alpha}_i(\tau) \mathcal{D} \psi^{\alpha *}_i(\tau)
\exp \left[  
-\int_0^{\beta}d\tau 
\right.\nonumber\\&& \left.
\left[ \sum_{i} \left\{ \psi^{\alpha *}_i(\tau) \left( \partial_{\tau}\psi^{\alpha}_i (\tau) \right)
-\left( \mu + i\sqrt{U} s_i(\tau) \right) \psi_i^{\alpha *} (\tau) \psi^{\alpha}_i (\tau) \right\} 
\right. \right.\nonumber\\&& \left.\left.
 -\frac{t}{Z}\sum_{\langle i,j \rangle}\left( 
 \psi_i^{\alpha *} (\tau) \psi^{\alpha}_j (\tau)
+ \psi_j^{\alpha *} (\tau) \psi^{\alpha}_i (\tau)
\right)\right] \right],
\end{eqnarray}
where $s_i(\tau)$ is an auxiliary field and the integral with respect to $s_i(\tau)$ is defined as
\begin{eqnarray}
\int\mathcal{D}s_i(\tau)&=&\prod_{i,\tau}\sqrt{\frac{N}{2\pi}}\int_{-\infty}^{\infty}ds_i(\tau).
\end{eqnarray}
In the large-$N$ limit, using saddle-point method, the auxiliary field $s_i(\tau)$ is 
replaced by $\overline{s}$ (see Ref. \cite{chen1998} and its references ), which makes the exponents of the partition function maximum.
Using Fourier transformation, we obtain
\begin{eqnarray}
Z_N&=& A e^{-\frac{N \beta L^d}{2} \overline{s}^2}\left[ Z_1\left(\overline{ s }\right) \right]^N,\\
Z_1\left( \overline{s}\right) &=&
\prod_{\bm k}\left(1- e^{-\beta \lambda_{\bm k}} \right)^{-1},\\
\lambda_{\bm k}&=& 
- \mu - i\sqrt{U} \overline{s}
-\frac{2t}{Z}\sum_{\delta=1}^{d} \cos k_\delta \label{eq:lambda}
\end{eqnarray}
where $A$ is a some real number caused by the fluctuation of $s_i\left( \tau \right)$ from $\overline{s}$, 
which does not contribute to following discussion
and the product of ${\bm k}$ runs over
the first Brillouin zone
${\bm k}=(2\pi /L) ( m_1,\cdots, m_d )$, with $m_i=1,2,\cdots,L$.
The stationary solution $\overline{s}$ must satisfy
\begin{eqnarray}
\frac{\partial}{\partial \overline{s}}\left[
-\frac{N \beta L^d}{2} \overline{s}^2 
- \sum_{\bm k} \ln \left(
1-e^{-\beta \lambda_{\bm k}}\right)\right]=0,
\end{eqnarray}
which yields,
\begin{eqnarray}
\overline{s}=\frac{i\sqrt{U}}{L^d}\sum_{\bm k}
\left[
  e^{\beta \lambda_{\bm k}}-1
\right]^{-1}.
\end{eqnarray}
The susceptibility $\chi$ is related to $\overline{s}$ by
\begin{eqnarray}
\chi \equiv \frac{1}{N}\int_{0}^{\beta}d\tau \sum_{i}\langle {\bm \psi}^{*}_i(\tau) \cdot {\bm \psi}_0(0) \rangle
= \left( -\mu-t-i\sqrt{U}\overline{s}\right)^{-1}.
\end{eqnarray}
Therefore, $\chi$ satisfies
\begin{eqnarray}
\chi^{-1}=-\mu-t+\frac{U}{L^d}\sum_{\bm k}\frac{1}{
\exp \left[ \frac{2\beta t}{Z}\sum_{\delta=1}^{d}\left( 1-\cos k_{\delta} \right) +\beta\chi^{-1} \right]-1}.
\label{eq:appchi}
\end{eqnarray}

\subsection{Derivation of Eq.(\ref{eq:app})}\label{app:detail_appr}
In Sec. \ref{sec:NinfFSS}, we derive 
$\chi = O\left((\beta t/Z)^{(2+d)/4}\right)$
by self-consistent analysis.
Namely, assuming the condition 
$\chi = O\left((\beta t/Z)^{(2+d)/4}\right)$, 
we prove the resulting solution satisfying this condition.
Here, assuming
\begin{eqnarray}
  (\beta t/Z)^{(2+d)/4}\chi^{-1}= O(1),\label{eq:assume1}
\end{eqnarray}
we provide an approximation form
\begin{eqnarray}
\sum_{\nu=1}^{\infty} e^{-\nu \beta \chi^{-1}} \left[ 
\sum_{n=1}^{L}\exp \left[ -\frac{2\nu \beta t}{Z} \left\{ 1-\cos\left(\frac{2\pi n}{L}\right)\right\}\right]
\right]^d \simeq \beta^{-1} \chi,\label{eq:appA}
\end{eqnarray}
which becomes exact under the condition that 
\begin{eqnarray}
d&>&2,\label{eq:assume2}\\
\beta t/L^2 &\ge& O(1), \label{eq:assume3}
\end{eqnarray}
and
\begin{eqnarray}
\beta t &\gg& 1.\label{eq:assume4}
\end{eqnarray}
To begin with, we rewrite the l.h.s. as
\begin{eqnarray}
\sum_{\nu=1}^{\infty} e^{-\nu \beta \chi^{-1}} \left[ 
\sum_{n=1}^{L}\exp \left[ -\frac{2\nu \beta t}{Z} \left\{ 1-\cos\left(\frac{2\pi n}{L}\right)\right\}\right]
\right]^d =
\sum_{\nu=1}^{\infty} e^{-\nu \beta \chi^{-1}} \left[1 +A_{\nu} \right]^d,\nonumber\\
=\sum_{\nu=1}^{\infty} e^{-\nu \beta \chi^{-1}}+
\sum_{\nu=1}^{\infty} e^{-\nu \beta \chi^{-1}} \left[ \sum_{\alpha=1}^d \frac{d!}{\alpha! \left(d-\alpha\right)!} A_\nu^{\alpha}\right],
\label{eq:aaa}
\end{eqnarray}
where
\begin{eqnarray}
A_\nu \equiv e^{-4\nu\beta t/Z} +2\sum_{n=1}^{-1+L/2} 
\exp \left[ -\frac{2\nu\beta t}{Z} \left\{ 1-\cos \left( \frac{2\pi n}{L}\right) \right\}\right].
\label{eq:aab}
\end{eqnarray}
Here we note that $\beta \chi^{-1} \simeq 0$.
(This is because $\beta\chi^{-1}=O((\beta t)^{-(d-2)/4})$ (by the condition Eq.(\ref{eq:assume1})) 
and by the condition Eq.(\ref{eq:assume2}) this is vanishing in the limit of Eq.(\ref{eq:assume4}).)
Since $\beta \chi^{-1} \simeq 0$,
the first term of the r.h.s. of Eq.(\ref{eq:aaa}) is approximated by the formula
\begin{eqnarray}
\sum_{\nu=1}^{\infty} e^{-\nu \beta \chi^{-1}}&=& \beta^{-1} \chi +O\left( (\beta \chi^{-1})^0 \right)
=O\left(\left(\beta t\right)^{\frac{d-2}{4}} \right).
\label{eq:aac}
\end{eqnarray}
Below we show that the second term of Eq.(\ref{eq:aaa}) is a correction term that vanishes in the limit of $\beta t \rightarrow \infty$. 
At first, $A_\nu$ is bounded as
\begin{eqnarray}
0<A_\nu &\leq& 
2\sum_{n=1}^{L/2} \exp\left[ - \frac{2\nu \beta t}{Z} \left\{ 1-\cos\left(\frac{2\pi n}{L}\right)\right\}\right],\nonumber\\
 &\leq& 2\int_0^{L/2}dp \exp\left[ - \frac{2\nu \beta t}{Z} \left\{ 1-\cos\left(\frac{2\pi p}{L}\right)\right\}\right]  ,\nonumber\\
 &\leq& 2\int_0^{L/2}dp \exp\left[ - \frac{2\nu \beta t}{Z} \left(\frac{8 p^2}{L^2} \right) \right]  ,\nonumber\\
 &\leq& \sqrt{\frac{\pi L^2 Z}{16\nu \beta t} },
\label{eq:aad}
\end{eqnarray}
Then, the second term of Eq. (\ref{eq:aaa}) is evaluated as
\begin{eqnarray}
0&<& \sum_{\alpha=1}^d \frac{d!}{\alpha! \left(d-\alpha\right)!} \left[\sum_{\nu=1}^{\infty} e^{-\nu \beta \chi^{-1}} A_\nu^{\alpha}\right],\nonumber\\
&\leq&  \sum_{\alpha=1}^d \frac{d!}{\alpha! \left(d-\alpha\right)!} \left[\sum_{\nu=1}^{\infty} e^{-\nu \beta \chi^{-1}} 
\left\{\frac{\pi L^2 Z}{16\nu \beta t} \right\}^{\alpha/2}  \right], \nonumber \\
 &\leq&  \sum_{\alpha=1}^d \frac{d!}{\alpha! \left(d-\alpha\right)!}
\left\{\frac{\pi L^2 Z}{16 \beta t} \right\}^{\alpha/2}  
\left[\int_{0}^{\infty}dp e^{-p \beta \chi^{-1}} p^{-\alpha/2} \right],\nonumber\\
&=&\sum_{\alpha=1}^d \frac{d!}{\alpha! \left(d-\alpha\right)!}
\left\{\frac{\pi Z}{16 (\beta t/L^2)} \right\}^{\alpha/2}  
\left[\int_{0}^{\infty}dq e^{-q} q^{-\alpha/2} \right] \left(\beta \chi^{-1}\right)^{\frac{\alpha-2}{2}}.
\end{eqnarray}
Since $\beta t/L^2\ge O(1)$, and $\beta \chi^{-1}\ll1$, the $\alpha=1$ term is dominant.
Therefore, the second term is of the same order as
$(\beta t/L^2)^{-1/2} (\beta\chi^{-1})^{-1/2}$.
By the condition Eq.(\ref{eq:assume1}), this is
$O\left( ( \beta t/L^2 )^{-1/2}\times (\beta t )^{(d-2)/8}\right)$.
Therefore, the ratio of the second and the first term of Eq.(\ref{eq:aaa}) becomes less than
$O\left( ( \beta t/L^2 )^{-1/2}\times (\beta t )^{-(d-2)/8}\right)$.
This is vanishing because of the condition Eqs. (\ref{eq:assume2}), (\ref{eq:assume3}) and (\ref{eq:assume4}).
Thus Eq.(\ref{eq:appA}) has been derived.

\section{Scaling function of superfluid density in large-N limit}\label{app:fssrhos}
In this section, we provide that the MFSS form of superfluid density using the $N$-component BH model.
The outline of this section is as follows.
First, we obtain the explicit definition of superfluid density,
which is estimated using the winding number in QMC simulation,
with an infinitesimal twist of phase of bosonic operator.
Next, we calculate the superfluid density of $N$-component BH model exactly.
The result reveals that the superfluid density $\rho_S$ is proportional to the susceptibility $\chi$.
Then, we derive the MFSS form of $\rho_S$ as that of $\chi$.

To start with, we derive an expression for the superfluid density $\rho_S$
introducing an infinitesimal twist of phase of bosonic operators.
Namely, we modify the Hamiltonian (\ref{eq:hamiltonian}) by
$b_{\alpha i}^{\dag}\rightarrow b_{\alpha i}^{\dag}e^{i\theta r_{i}^{z}} $,
$b_{\alpha i}\rightarrow b_{\alpha i}e^{-i\theta r_{i}^{z}} $,
where $r_{i}^{z}$ is the $z$-coordinate of the site $i$.
(Because of the periodic boundary condition,
$\theta$ should be discrete. That is, $\theta=2\pi n/L$ where $n$ is integer.
However, considering a sufficiently large system, we regard $\theta$ as 
a continuous real number.)
Then, we define the twisted Hamiltonian $\mathcal{H}_{N\theta}$, the partition function $Z_{N\theta}$ and the free energy $F_{N\theta}$ as,
\begin{eqnarray}
{\mathcal H}_{N\theta}&=&-\frac{t}{Z}\sum_{\alpha=1}^{N}\sum_{\langle i,j \rangle}\left( 
b^{\dag}_{\alpha i} b_{\alpha j} e^{i\theta\left( r_{i}^{z}-r_{j}^{z} \right)}
+b_{\alpha i} b^{\dag}_{\alpha j} e^{-i\theta\left( r_{i}^{z}-r_{j}^{z} \right)}
\right)
-\mu \sum_{\alpha=1}^{N}\sum_{i} b^{\dag}_{\alpha i} b_{\alpha i}\nonumber\\
&&+\frac{U}{2N} \sum_{\alpha=1}^{N}\sum_{\beta=1}^{N} \sum_{i} b^{\dag}_{\alpha i} b^{\dag}_{\beta i} b_{\beta i} b_{\alpha i},
\label{eq:ham_Ntheta}\\
Z_{N\theta}&=&{\rm Tr}\left[ e^{-\beta {\mathcal{H}}_{N\theta}}\right],\label{eq:Z_Ntheta}\\
F_{N\theta}&=&-\frac{1}{N}\ln Z_{N\theta}. \label{eq:F_Ntheta}
\end{eqnarray}
The superfluid density $\rho_S$ is defined with this twisted free energy $F_{N\theta}$ as,
\begin{eqnarray}
\rho_S \equiv \frac{1}{2\beta L^{d} \left(t/Z\right)} \left.\frac{\partial^2 F_{N\theta}}{\partial \theta^2}\right|_{\theta=0}.
\label{eq:rhos}
\end{eqnarray}
Next, we calculate the $\rho_S$ in the large-$N$ limit.
The partition function $Z_{N\theta}$ is obtained as well as 
the non-twisted partition function (See Appendix. \ref{app:detail_self}) as,
\begin{eqnarray}
Z_{N\theta}&=& A e^{-\frac{N \beta L^d}{2} \overline{s}^2}\left[ Z_{1\theta}\left(\overline{ s }\right) \right]^N,\\
Z_{1\theta}\left( \overline{s}\right) &=&
\prod_{\bm k}\left(1- e^{-\beta \lambda_{{\bm k}\theta}} \right)^{-1},\\
\lambda_{{\bm k}\theta}&=& 
- \mu - i\sqrt{U} \overline{s}
-\frac{2t}{Z}\sum_{\delta=1}^{d-1} \cos k_\delta
-\frac{2t}{Z} \cos \left(k_z +\theta \right).
\end{eqnarray}
The derivation of the free energy is straightforward using this partition function.
Then, we obtain the superfluid density is
\begin{eqnarray}
\rho_S=\frac{1}{L^d}\sum_{\bm k}\frac{\cos k_z}{e^{\beta \lambda_{\bm k}}-1},
\end{eqnarray}
with $\lambda_{\bm k}$ defined in Eq. (\ref{eq:lambda}).
This superfluid density is smaller than the total density of particle,
\begin{eqnarray}
\rho&\equiv&-\frac{1}{L^{d}\beta}\frac{\partial F_{N{\theta=0}}}{\partial \mu},\nonumber\\
&=&\frac{1}{L^d}\sum_{\bm k}\frac{1}{e^{\beta\lambda_{\bm k}}-1},
\end{eqnarray}
and larger than the density of particles of ${\bm k}=0$,
\begin{eqnarray}
\rho_0=\frac{1}{L^d}\frac{1}{e^{\beta\chi^{-1}}-1}.
\end{eqnarray}
That is,
\begin{eqnarray}
\rho_0\leq \rho_S\leq \rho. \label{eq:n0_rhos_n}
\end{eqnarray}
As shown in Appendix \ref{app:detail_appr},
\begin{eqnarray}
\rho = \rho_0 = \frac{\chi}{L^d\beta},
\end{eqnarray}
under the condition $\beta t/L^2 \geq O(1)$, $d>2$ and $\beta t \rightarrow \infty$.
Using the inequality (\ref{eq:n0_rhos_n}),
we obtain 
\begin{eqnarray}
\rho_S = \frac{\chi}{L^d\beta}.
\end{eqnarray}
As shown in Sec. \ref{sec:NinfFSS}, we derive the MFSS form of $\rho_S$ as
\begin{eqnarray}
\rho_S &=& L^{-\frac{d+2}{2}}P^{UZ/t}_{\rho_S}\left(L^{\frac{d+2}{2}} Z\left(-\frac{\mu}{t}-1\right), \frac{\beta t}{L^2Z} \right), \\
P^{u}_{\rho_S}\left(x,y\right)&\equiv& \frac{2 y^{-1}}{x+\sqrt{x^2+4uy^{-1}}}.
\end{eqnarray}
The applicability condition of this MFSS form is $d>2$ and $\beta t/L^2 = O(1)$.

\newpage 

\begin{thebibliography}{21}
\expandafter\ifx\csname natexlab\endcsname\relax\def\natexlab#1{#1}\fi
\expandafter\ifx\csname bibnamefont\endcsname\relax
  \def\bibnamefont#1{#1}\fi
\expandafter\ifx\csname bibfnamefont\endcsname\relax
  \def\bibfnamefont#1{#1}\fi
\expandafter\ifx\csname citenamefont\endcsname\relax
  \def\citenamefont#1{#1}\fi
\expandafter\ifx\csname url\endcsname\relax
  \def\url#1{\texttt{#1}}\fi
\expandafter\ifx\csname urlprefix\endcsname\relax\def\urlprefix{URL }\fi
\providecommand{\bibinfo}[2]{#2}
\providecommand{\eprint}[2][]{\url{#2}}

\bibitem[{\citenamefont{Greiner et~al.}(2002)\citenamefont{Greiner, Mandel,
  Esslinger, H{\"a}nsch, and Bloch}}]{greiner2002}
\bibinfo{author}{\bibfnamefont{M.}~\bibnamefont{Greiner}},
  \bibinfo{author}{\bibfnamefont{O.}~\bibnamefont{Mandel}},
  \bibinfo{author}{\bibfnamefont{T.}~\bibnamefont{Esslinger}},
  \bibinfo{author}{\bibfnamefont{T.~W.} \bibnamefont{H{\"a}nsch}},
  \bibnamefont{and} \bibinfo{author}{\bibfnamefont{I.}~\bibnamefont{Bloch}},
  \bibinfo{journal}{Nature} \textbf{\bibinfo{volume}{415}}, \bibinfo{pages}{39}
  (\bibinfo{year}{2002}).

\bibitem[{\citenamefont{Kato et~al.}(2008)\citenamefont{Kato, Zhou, Kawashima,
  and Trivedi}}]{kato2008}
\bibinfo{author}{\bibfnamefont{Y.}~\bibnamefont{Kato}},
  \bibinfo{author}{\bibfnamefont{Q.}~\bibnamefont{Zhou}},
  \bibinfo{author}{\bibfnamefont{N.}~\bibnamefont{Kawashima}},
  \bibnamefont{and} \bibinfo{author}{\bibfnamefont{N.}~\bibnamefont{Trivedi}},
  \bibinfo{journal}{Nat. Phys.} \textbf{\bibinfo{volume}{4}},
  \bibinfo{pages}{617} (\bibinfo{year}{2008}).

\bibitem[{\citenamefont{Jaksch et~al.}(1998)\citenamefont{Jaksch, Bruder,
  Cirac, Gardiner, and Zoller}}]{jaksch1998}
\bibinfo{author}{\bibfnamefont{D.}~\bibnamefont{Jaksch}},
  \bibinfo{author}{\bibfnamefont{C.}~\bibnamefont{Bruder}},
  \bibinfo{author}{\bibfnamefont{J.~I.} \bibnamefont{Cirac}},
  \bibinfo{author}{\bibfnamefont{C.~W.} \bibnamefont{Gardiner}},
  \bibnamefont{and} \bibinfo{author}{\bibfnamefont{P.}~\bibnamefont{Zoller}},
  \bibinfo{journal}{Phys. Rev. Lett.} \textbf{\bibinfo{volume}{81}},
  \bibinfo{pages}{3108} (\bibinfo{year}{1998}).

\bibitem[{\citenamefont{Fisher et~al.}(1989)\citenamefont{Fisher, Weichman,
  Grinstein, and Fisher}}]{fisher1989}
\bibinfo{author}{\bibfnamefont{M.~P.~A.} \bibnamefont{Fisher}},
  \bibinfo{author}{\bibfnamefont{P.~B.} \bibnamefont{Weichman}},
  \bibinfo{author}{\bibfnamefont{G.}~\bibnamefont{Grinstein}},
  \bibnamefont{and} \bibinfo{author}{\bibfnamefont{D.~S.}
  \bibnamefont{Fisher}}, \bibinfo{journal}{Phys.\ Rev.\ B}
  \textbf{\bibinfo{volume}{40}}, \bibinfo{pages}{546} (\bibinfo{year}{1989}).

\bibitem[{\citenamefont{Capogrosso-Sansone
  et~al.}(2007)\citenamefont{Capogrosso-Sansone, Prokof'ev, and
  Svistunov}}]{capogrosso2007}
\bibinfo{author}{\bibfnamefont{B.}~\bibnamefont{Capogrosso-Sansone}},
  \bibinfo{author}{\bibfnamefont{N.~V.} \bibnamefont{Prokof'ev}},
  \bibnamefont{and} \bibinfo{author}{\bibfnamefont{B.~V.}
  \bibnamefont{Svistunov}}, \bibinfo{journal}{Phys.\ Rev.\ B}
  \textbf{\bibinfo{volume}{75}}, \bibinfo{pages}{134302}
  (\bibinfo{year}{2007}).

\bibitem[{\citenamefont{Kawashima and Kato}(2009)}]{kawashima2009}
\bibinfo{author}{\bibfnamefont{N.}~\bibnamefont{Kawashima}} \bibnamefont{and}
  \bibinfo{author}{\bibfnamefont{Y.}~\bibnamefont{Kato}},
  \bibinfo{journal}{Journal of Physics: Conference Series}
  \textbf{\bibinfo{volume}{143}}, \bibinfo{pages}{012012}
  (\bibinfo{year}{2009}).

\bibitem[{\citenamefont{Binder et~al.}(1985)\citenamefont{Binder, Nauenberg,
  Privman, and Young}}]{binder1985}
\bibinfo{author}{\bibfnamefont{K.}~\bibnamefont{Binder}},
  \bibinfo{author}{\bibfnamefont{M.}~\bibnamefont{Nauenberg}},
  \bibinfo{author}{\bibfnamefont{V.}~\bibnamefont{Privman}}, \bibnamefont{and}
  \bibinfo{author}{\bibfnamefont{A.~P.} \bibnamefont{Young}},
  \bibinfo{journal}{Phys.\ Rev.\ B} \textbf{\bibinfo{volume}{31}},
  \bibinfo{pages}{1498} (\bibinfo{year}{1985}).

\bibitem[{\citenamefont{Brankov et~al.}(2000)\citenamefont{Brankov, Danchev,
  and Tonchev}}]{brankov}
\bibinfo{author}{\bibfnamefont{J.~G.} \bibnamefont{Brankov}},
  \bibinfo{author}{\bibfnamefont{D.~M.} \bibnamefont{Danchev}},
  \bibnamefont{and} \bibinfo{author}{\bibfnamefont{N.~S.}
  \bibnamefont{Tonchev}}, \emph{\bibinfo{title}{THEORY OF CRITICAL PHENOMENA IN
  FINITE-SIZE SYSTEMS Scaling and Quantum Effects}} (\bibinfo{publisher}{World
  Scientific}, \bibinfo{year}{2000}), chap.~\bibinfo{chapter}{6}.

\bibitem[{\citenamefont{Luijten et~al.}(1999)\citenamefont{Luijten, Binder, and
  Bl{\" o}te}}]{luijten1999}
\bibinfo{author}{\bibfnamefont{E.}~\bibnamefont{Luijten}},
  \bibinfo{author}{\bibfnamefont{K.}~\bibnamefont{Binder}}, \bibnamefont{and}
  \bibinfo{author}{\bibfnamefont{H.~W.~J.} \bibnamefont{Bl{\" o}te}},
  \bibinfo{journal}{Eur.\ Phys.\ J.\ B} \textbf{\bibinfo{volume}{9}},
  \bibinfo{pages}{289} (\bibinfo{year}{1999}).

\bibitem[{\citenamefont{Jones and Young}(2005)}]{jones2005}
\bibinfo{author}{\bibfnamefont{J.~L.} \bibnamefont{Jones}} \bibnamefont{and}
  \bibinfo{author}{\bibfnamefont{A.~P.} \bibnamefont{Young}},
  \bibinfo{journal}{Phys.\ Rev.\ B} \textbf{\bibinfo{volume}{71}},
  \bibinfo{pages}{174438} (\bibinfo{year}{2005}).

\bibitem[{\citenamefont{Singh and Pathria}(1988)}]{singh1988}
\bibinfo{author}{\bibfnamefont{S.}~\bibnamefont{Singh}} \bibnamefont{and}
  \bibinfo{author}{\bibfnamefont{R.~K.} \bibnamefont{Pathria}},
  \bibinfo{journal}{Phys.\ Rev.\ B} \textbf{\bibinfo{volume}{38}},
  \bibinfo{pages}{2740} (\bibinfo{year}{1988}).

\bibitem[{\citenamefont{Chen and Dohm}(1998)}]{chen1998}
\bibinfo{author}{\bibfnamefont{X.~S.} \bibnamefont{Chen}} \bibnamefont{and}
  \bibinfo{author}{\bibfnamefont{V.}~\bibnamefont{Dohm}},
  \bibinfo{journal}{Eur.\ Phys.\ J.\ B} \textbf{\bibinfo{volume}{5}},
  \bibinfo{pages}{529} (\bibinfo{year}{1998}).

\bibitem[{\citenamefont{{\v S}makov and S{\o}rensen}(2005)}]{smakov2005}
\bibinfo{author}{\bibfnamefont{J.}~\bibnamefont{{\v S}makov}} \bibnamefont{and}
  \bibinfo{author}{\bibfnamefont{E.}~\bibnamefont{S{\o}rensen}},
  \bibinfo{journal}{Phys.\ Rev.\ Lett.} \textbf{\bibinfo{volume}{95}},
  \bibinfo{pages}{180603} (\bibinfo{year}{2005}).

\bibitem[{\citenamefont{Zhao et~al.}()\citenamefont{Zhao, Sandvik, and
  Ueda}}]{zhao2008}
\bibinfo{author}{\bibfnamefont{J.}~\bibnamefont{Zhao}},
  \bibinfo{author}{\bibfnamefont{A.~W.} \bibnamefont{Sandvik}},
  \bibnamefont{and} \bibinfo{author}{\bibfnamefont{K.}~\bibnamefont{Ueda}},
  \bibinfo{note}{arXiv:0806.3603}.

\bibitem[{\citenamefont{Binder and Wang}(1989)}]{binder1989}
\bibinfo{author}{\bibfnamefont{K.}~\bibnamefont{Binder}} \bibnamefont{and}
  \bibinfo{author}{\bibfnamefont{J.-S.} \bibnamefont{Wang}},
  \bibinfo{journal}{J. Stat. Phys.} \textbf{\bibinfo{volume}{55}},
  \bibinfo{pages}{87} (\bibinfo{year}{1989}).

\bibitem[{\citenamefont{Zannetti}(1980)}]{zannetti1980}
\bibinfo{author}{\bibfnamefont{M.}~\bibnamefont{Zannetti}},
  \bibinfo{journal}{Phys.\ Rev.\ B} \textbf{\bibinfo{volume}{22}},
  \bibinfo{pages}{5267} (\bibinfo{year}{1980}).

\bibitem[{\citenamefont{Cesare}(1982)}]{cesare1982}
\bibinfo{author}{\bibfnamefont{L.~D.} \bibnamefont{Cesare}},
  \bibinfo{journal}{Il Nuovo Cim.\ D} \textbf{\bibinfo{volume}{1}},
  \bibinfo{pages}{289} (\bibinfo{year}{1982}).

\bibitem[{\citenamefont{Kato et~al.}(2007)\citenamefont{Kato, Suzuki, and
  Kawashima}}]{kato2007}
\bibinfo{author}{\bibfnamefont{Y.}~\bibnamefont{Kato}},
  \bibinfo{author}{\bibfnamefont{T.}~\bibnamefont{Suzuki}}, \bibnamefont{and}
  \bibinfo{author}{\bibfnamefont{N.}~\bibnamefont{Kawashima}},
  \bibinfo{journal}{Phys.\ Rev.\ E} \textbf{\bibinfo{volume}{75}},
  \bibinfo{pages}{066703} (\bibinfo{year}{2007}).

\bibitem[{\citenamefont{Kato and Kawashima}(2009)}]{kato2009}
\bibinfo{author}{\bibfnamefont{Y.}~\bibnamefont{Kato}} \bibnamefont{and}
  \bibinfo{author}{\bibfnamefont{N.}~\bibnamefont{Kawashima}},
  \bibinfo{journal}{Phys.\ Rev.\ E} \textbf{\bibinfo{volume}{79}},
  \bibinfo{pages}{021104} (\bibinfo{year}{2009}).

\bibitem[{\citenamefont{Pollock and Ceperley}(1987)}]{pollock1987}
\bibinfo{author}{\bibfnamefont{E.~L.} \bibnamefont{Pollock}} \bibnamefont{and}
  \bibinfo{author}{\bibfnamefont{D.~M.} \bibnamefont{Ceperley}},
  \bibinfo{journal}{Phys.\ Rev.\ B} \textbf{\bibinfo{volume}{36}},
  \bibinfo{pages}{8343} (\bibinfo{year}{1987}).

\bibitem[{\citenamefont{Singh and Pathria}(1986)}]{singh1986}
\bibinfo{author}{\bibfnamefont{S.}~\bibnamefont{Singh}} \bibnamefont{and}
  \bibinfo{author}{\bibfnamefont{R.~K.} \bibnamefont{Pathria}},
  \bibinfo{journal}{Phys.\ Rev.\ B} \textbf{\bibinfo{volume}{34}},
  \bibinfo{pages}{2045} (\bibinfo{year}{1986}).

\end{thebibliography}

\end{document}